\documentclass[sigconf, 9pt]{acmart}

\usepackage{soul}         %
\usepackage{xcolor}       %
\usepackage{etoolbox}     %

\sethlcolor{green!20}     %

\newcommand{\del}[2][]{}

\definecolor{eidhl}{RGB}{176,84,0}

\usepackage{float}
\usepackage{listings}
\usepackage{newfloat}
\usepackage{makecell}
\usepackage{subcaption}

\DeclareCaptionStyle{ruled}{labelfont=normalfont,labelsep=colon,strut=off}
\lstdefinestyle{eidexample}{moredelim=**[is][\color{eidhl}\bfseries]{@@}{@@},
  basicstyle={\scriptsize\ttfamily}}
\floatstyle{ruled}
\newfloat{listing}{tb}{lst}
\floatname{listing}{Listing}

\newcommand{\yc}[1]{\ignorespaces}

\setcopyright{cc}
\setcctype{by}

\renewcommand\footnotetextcopyrightpermission[1]{}



\title{Bridging the Gap: A Longitudinal Analysis of Extended Identifiers in the Post-Cookie Era}

\author{Michael Smith}
\email{ms255@iu.edu}
\affiliation{%
  \institution{Indiana University Bloomington}
  \city{Bloomington}
  \state{Indiana}
  \country{USA}
}

\author{Riley Grossman}
\email{rag24@njit.edu}
\affiliation{%
  \institution{New Jersey Institute of Technology}
  \city{Newark}
  \state{New Jersey}
  \country{USA}
}

\author{Krzysztof Franaszek}
\email{krzysztof@adalytics.io}
\affiliation{%
  \institution{Adalytics}
  \city{New York}
  \state{New York}
  \country{USA}
}

\author{Antonio Torres-Ag\"{u}ero}
\email{antonio@deepsee.io}
\affiliation{%
  \institution{DeepSee.io}
  \city{San Francisco}
  \state{California}
  \country{USA}
}

\author{Pritam Sen}
\email{ps37@njit.edu}
\affiliation{%
  \institution{New Jersey Institute of Technology}
  \city{Newark}
  \state{New Jersey}
  \country{USA}
}

\author{Cristian Borcea}
\email{borcea@njit.edu}
\affiliation{%
  \institution{New Jersey Institute of Technology}
  \city{Newark}
  \state{New Jersey}
  \country{USA}
}

\author{Yi Chen}
\email{yi.chen@njit.edu}
\affiliation{%
  \institution{New Jersey Institute of Technology}
  \city{Newark}
  \state{New Jersey}
  \country{USA}
}

\begin{abstract}
As third-party cookies fade because of browser restrictions, the online advertising ecosystem is turning to extended identifiers (EIDs) as an alternative. EIDs are persistent user identifiers, such as hashed email addresses, that are employed to link users across domains and devices. This paper presents a 41-month longitudinal study examining EID usage in over 145 million HTTP header bidding requests sent to six major supply-side platforms (SSPs) from 616,539 websites. %
Our findings show that EIDs are widely used and are becoming increasingly prevalent in the digital advertising ecosystem, reaching 83.76\% of studied websites by May 2025. %
Our analysis of the 18 popular EID providers that account for 99.42\% of all transmitted EIDs in our dataset raises concerns about the readiness of EIDs as an alternative to third-party cookie tracking. In terms of accuracy, only one identity provider consistently recognizes and identifies that the visitor is a self-identified bot crawler, and many providers regularly transmit multiple EIDs for the same visitor. We also identify privacy concerns with EIDs, as 12 of the providers create persistent EIDs that can identify the same user across visits, websites, devices, and months. Finally, we found that 16 providers transmit EIDs on EU websites without user consent. 

\end{abstract}

\begin{document}
\maketitle
\pagestyle{plain}

\section{Introduction}

Third-party cookies have been essential to real-time bidding (RTB), a \$22.3 billion market~\cite{market_size} and a crucial component of the larger digital advertising industry. In RTB, publishers make their ad spaces available for each individual visitor. Ad tech companies, such as supply-side platforms (SSPs) and demand-side platforms (DSPs), representing publishers and advertisers, respectively, conduct automated, real-time auctions to decide which ad is shown when a page loads. The auction allows advertisers to decide exactly how much a given impression is worth to them, and ensures publishers gain the maximum ad revenue for a given impression. Cookies help RTB by enabling the persistent identification of a user’s browser, so information such as browsing history, preferences, and inferred interests can be collected. The collected user profiles enable advertisers to more accurately estimate the value of showing an ad to the user, and thus, increase bid values and publishers’ ad revenues in a process sometimes called targeted advertising. 

\begin{figure}[t!]
  \centering
  \includegraphics[width=0.80\linewidth]{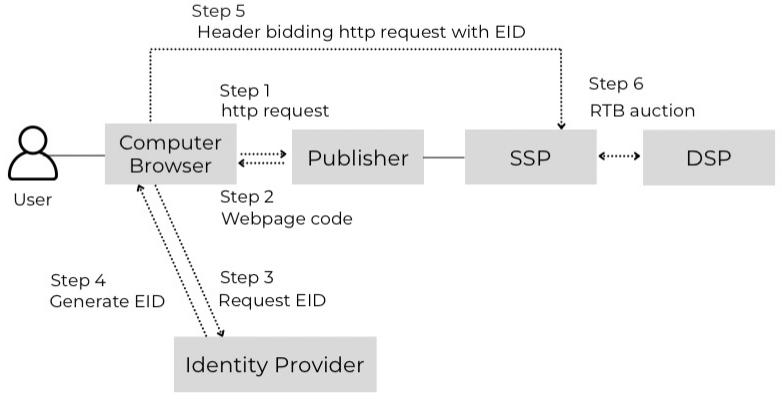}
  \caption{Workflow illustrating the EID process}
  \label{fig:eid_workflow}
\end{figure}

Due to privacy concerns, several companies (e.g., Apple, Microsoft, and Mozilla) have implemented policies in the last several years to block third-party cookies by default~\cite{shankar2023,munir2023cookiegraph,chen2021cookieswap}. During this period, Google announced plans to deprecate third-party cookies by 2022. However, these plans were then delayed several times and ultimately abandoned~\cite{roth2025,drum_google_cookies,adexchanger2024sandbox}. 

Partially in response to the anticipated deprecation of third-party cookies, advertising technology companies introduced extended identifiers (EIDs) as a third-party cookie alternative. EIDs typically rely on first-party data, such as the hashed email address provided at login, to create a persistent user identifier~\cite{iiq_EIDs}. EIDs are promoted by the advertising industry as a privacy-preserving alternative to third-party cookies~\cite{admonsters_eids_10, iiq_EIDs, id5id, audigent2023hadron,sivan2025} because the EID is encrypted and, thus, only accessible to those with access granted by the identity providers. Centralizing control over who can decode EIDs lets identity providers release user identities according to user preferences and privacy regulations~\cite{id5id, iiq_EIDs, audigent2023hadron, admonsters_eids_10, sivan2025}. In contrast, third-party cookies, managed by companies that often ignore users’ privacy preferences, are set in the browser and are accessible to many third parties through cookie syncs.

Figure~\ref{fig:eid_workflow} shows how EIDs are used in header bidding (i.e., an implementation of RTB that allows multiple SSPs to auction off the impression at once~\cite{pachilakis19}). After a user requests the publisher’s content via an HTTP request (Step 1), the publisher’s webpage, which may contain advertising scripts, is returned (Step 2). If an identity provider’s script is embedded in the webpage, it processes the available identity signals and sends them to the identity provider (Step 3). A resolved EID is sent back to the browser (Step 4).
When the visitor is logged in, the identity provider can deterministically match the visit to a known identity (e.g., email address). When no such signal is available, the provider probabilistically derives the EID from what it can observe, such as the IP address, first-party cookies, and device characteristics (e.g., screen resolution, operating system, browser, and user-agent string). During client-side header bidding auctions, the EID is sent in an HTTP request to the SSP (Step 5), which then conducts an RTB auction among DSPs (Step 6). Including the EID in the bid request allows advertisers to identify and target specific (groups of) users without cookies.

In addition to serving as third-party cookie replacements, EIDs enable cross-device and cross-browser tracking. This process, known as ID bridging, involves linking different identifiers (e.g., cookies, hashed emails, or browser fingerprints~\cite{liu2025fingerprinting}) that represent the same user in varying contexts (e.g., a person’s laptop and mobile device). Deterministic methods include linking different browsing sessions based on a unique identifier, such as the email address used to log in to a user account. Probabilistic ID bridging techniques are similar to a traditional tracking technique known as device fingerprinting, which uses device or browser characteristics such as screen resolution and installed browser extensions to identify multiple visits from the same device/browser without third-party cookies~\cite{tiwari2023fingerprinting}. The potential for cross-device tracking and the reliance on probabilistic ID bridging techniques have introduced concerns into the accuracy and privacy-preserving qualities of EIDs~\cite{iiq_EIDs,sevilla2022, sivan2025}. %

These developments raise two research questions: (1) how prevalent have EIDs become in response to third-party cookie deprecation announcements?, (2) How are EIDs typically transmitted, and is this process aligned with the goal of becoming a privacy-preserving and third-party cookie alternative? %

This paper addresses these questions through a 41-month longitudinal empirical study. We analyzed a subset of the HTTP Archive dataset containing over 145 million header bidding HTTP requests between crawled websites and six SSPs—Magnite, Yieldmo, Sonobi, Amazon Ads, Index Exchange, and GumGum. This dataset allowed us to examine the use of EIDs for header bidding across multiple webpages from 616,539 websites on both mobile and desktop devices, covering the period from January 2022 to May 2025. The crawler never signs in, so every EID we observe is derived from observable signals rather than a logged-in identity.

The analysis of this dataset in Section~\ref{subsec:prevalence} enabled us to make two major findings pertaining to the prevalence of EIDs in header bidding. First, EIDs are becoming increasingly popular, as the percentage of visited websites transmitting an EID increased from 28.92\% in January 2022 to 83.76\% in May 2025. Second, although early increases in prevalence may have been related to Google's third-party cookie deprecation announcements, this is no longer the case: prevalence rose substantially after Google abandoned those plans in July 2024~\cite{drum_google_cookies}.

Our analyses of the EID sharing behaviors of 18 popular identity providers uncovered concerns about their accuracy and privacy-preserving qualities. First, we find that all but one identity provider demonstrates inconsistent behavior and sometimes transmits multiple EIDs for a single visitor (see Section~\ref{subsub:multiple}). This behavior is particularly concerning in cases where SSPs receive different user identities from the identity provider as it may lead to unfair auctions where advertisers have different beliefs about the user's identity and interests. Second, only one of the providers correctly identifies the crawler as bot traffic despite the crawler self-identifying (see Section~\ref{subsub:bots}). Thus, advertisers who bid based on the EIDs transmitted by the other 17 providers may waste money to show their ads to a bot. Third, 12 of the 18 providers generate persistent EIDs that can identify a user across websites, devices, and months (see Section~\ref{subsub:reuse}). In addition to obvious privacy concerns, this raises accuracy concerns as the collected data represents a scenario where identity providers do not have useful signals of user identity. Finally, we find that 16 of the providers transmit EIDs on websites that appear to operate in the European Union (EU) (see Section~\ref{subsub:consent}). This is concerning as explicit consent was not granted by the user as required by the General Data Protection Regulation (GDPR).
To facilitate reproducibility, the processed dataset and code are available at \url{https://github.com/rag24/longitudinal_eid_study}.

\vspace{-0.2in}
\section{Related Work}
\label{sec:background_rw}

Prior research has qualitatively compared EIDs to cookies, and discussed the implications for Web privacy~\cite{sivan2025,sevilla2022}. We are unaware of prior academic research on EID adoption. Industry reports have suggested a growing number of alternative IDs, including EIDs, but information on their adoption rates is scarce and proprietary~\cite{industry_id5_report}. Furthermore, we believe we are the first to identify potential issues with the current behavior of EID providers that question the readiness of EIDs as a privacy-preserving identifier.
Next, we discuss tracking technologies related to EIDs.

\textbf{Cross-Device Tracking.} Several papers demonstrate the viability of probabilistically matching users across devices based on commonly disclosed information such as IP address, first/third-party cookies, and user accounts~\cite{diaz2015crossdevice,brookman2017crossdevice,zimmeck2017crossdevice}. A more recent paper found that inferred user interests based on interactions with Amazon Alexa devices were used to serve targeted advertisements on other devices using the same Amazon account as Alexa~\cite{iqbal2023alexa}. These demonstrated ID bridging techniques are likely similar to the proprietary ID bridging methodologies of EID providers, and demonstrate the feasibility of cross-device tracking. 

\textbf{Fingerprinting.} Some identity providers also use fingerprinting to generate EIDs. This probabilistic ID bridging technique infers user identities when no deterministic signals are available (e.g., not logged in). Prior research has developed techniques for detecting fingerprinting scripts and measured their prevalence on the Web~\cite{iqbal2021fingerprinters,englehardt2016online,luo25}, identified privacy concerns~\cite{tiwari2023fingerprinting}, and shown that fingerprints identify users and serve targeted advertisements~\cite{liu2025fingerprinting}. 

\textbf{First-Party Cookies.} First-party cookies (i.e., set by the visited domain and accessible only to it) are one identity signal used by EID providers. Originally assumed useful only for same-site tracking, prior research showed they also enable tracking when third-party trackers use CNAME cloaking to reach them or are embedded in a first-party context~\cite{bahrami25,chen2021cookieswap}. First-party cookies and fingerprints are two examples of \emph{identity signals} that a provider may utilize to generate an EID when a logged-in identity is not available.

\section{Method}
\label{sec:method}

\begin{sloppypar}
We investigate the real-world adoption of EIDs, based on HTTP requests containing header bidding information collected and stored by the open-source HTTP Archive~\cite{httparchive}. Header bidding is a programmatic advertising mechanism in which publishers concurrently solicit bids from multiple demand sources (i.e., advertisers or DSPs)~\cite{pachilakis19}. Since the identity of the user who will be served the advertisement is an important component of bid prices, any user identifiers that are available (including EIDs) are included in the header bidding requests. Thus, we leverage these requests to study the prevalence and characteristics of EID sharing. %
\end{sloppypar}

\textbf{Data Collection and Scope.} Monthly, the HTTP Archive conducts a massive web crawl from automated Chrome browsers. The crawler visits, but does not interact with, webpages from over 16 million websites. For each website, it visits the home page and one secondary page (selected using the Chrome UX Report (CrUX)).
The list is drawn from CrUX, so it covers the websites most visited by real Chrome users in each year, but this also means we lack a consistent sample of websites over time. To address this, we repeat our prevalence analyses in Appendix~\ref{app:subset} and show this does not drive our findings of increasing EID usage. Each of the targeted webpages is visited from an Android smartphone (i.e., mobile device) and a Linux virtual machine (i.e., desktop) from IP addresses provided by Google Cloud and located within the United States. Thus, there are four different crawls to a single website per month. The crawler is built on top of the open-source WebPageTest~\cite{webtest} framework to collect HTTP requests generated from the webpage visit. As a result, the crawler self-declares as a bot by adding ``PTST'' to the User-Agent string. 

We query the HTTP Archive's dataset using Google's BigQuery to collect the HTTP requests from January 2022 to May 2025 between the visited websites and the header bidding endpoints of six widely used SSPs: Magnite (formerly Rubicon), Amazon Ads, Index Exchange (formerly Casale), Sonobi, GumGum, and Yieldmo. These six SSPs were chosen because there is a high volume of header bidding auctions including them in the HTTP Archive dataset. Although EIDs may be communicated during either RTB or header bidding requests, we collect header bidding requests because the RTB requests are sent server-side between SSPs and DSPs and are not visible to the crawler. EIDs from header bidding requests are visible to the crawler because these requests are sent client-side. %

\textbf{EID Extraction.} The header bidding requests and EIDs contained within them can be formatted differently, depending on the SSP. Prebid is an open-source framework to standardize header bidding auctions. SSPs that utilize Prebid provide a Prebid adapter that states how to format the header bidding request for that particular SSP~\cite{prebid_github}. We manually evaluate HTTP header bidding requests and utilize the available Prebid adapters for each SSP to determine how they format their header bidding requests and store EIDs. 
After identifying each SSP's header bidding request format, we implement Python code that processes each HTTP request to determine whether it contains EIDs, and stores their number, values, and generating identity provider.

Within a request, the EIDs for a visit are transmitted together in a way that allows downstream parties to match each EID with its provider. Locating the list of EIDs is dependent on the SSP, since each chooses how to encode their incoming requests. We resolve this per SSP using the published Prebid adapter code~\cite{prebid_github} and manual inspection of raw requests. Appendix~\ref{app:extraction} shows an example request to each SSP with the EIDs highlighted.

\begin{figure*}[t!]
  \centering
  \includegraphics[width=0.85\linewidth]{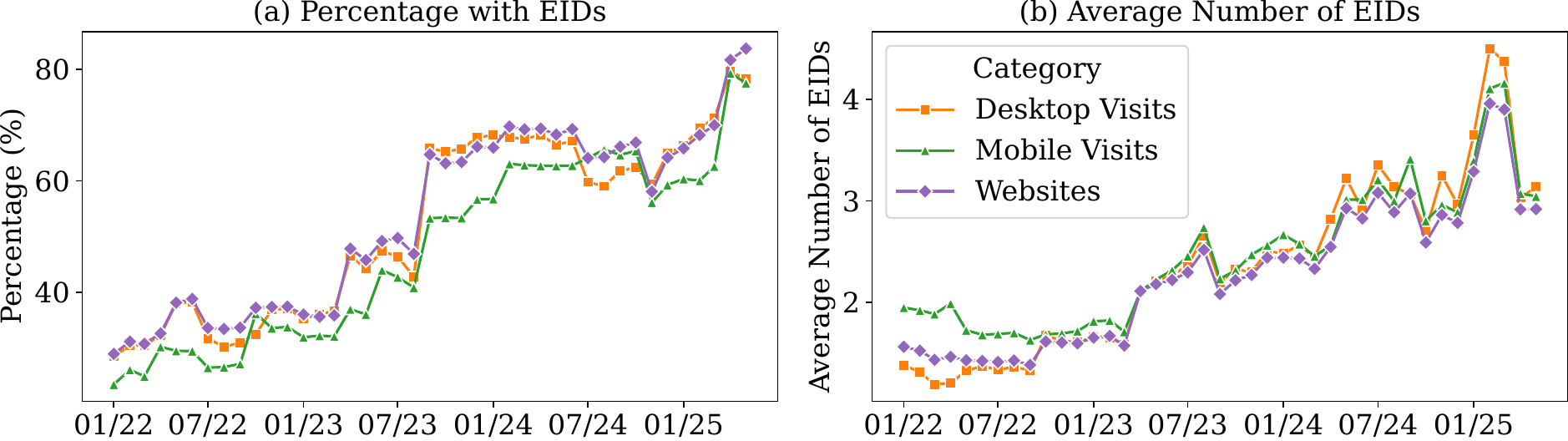}
  \caption{Prevalence of EIDs Over Time}
  \label{fig:prevalence_eids_time}
\end{figure*}

\textbf{Request Aggregation.} A single crawler visit may result in HTTP requests between the website and up to all six of the SSPs (depending on partnerships). Due to multiple ad slots per page, a visit may trigger multiple HTTP requests between the website and each SSP. %
Many of these HTTP requests contain redundant information (e.g., the same EID), so analysis at the request level would count a single webpage visit's EID transmission many times. We therefore combine all HTTP requests sent to each SSP and record every unique combination of EID provider and value, yielding one record per webpage visit, EID, and SSP. We retain null EID values (e.g., ``0'') since the website still integrated an EID module and attempted to resolve the visitor's identity.

\section{Results and Analysis}
\label{sec:results}

We retrieved 145,363,816 HTTP requests between the websites visited by the HTTP Archive's crawls and the 6 SSPs. After processing and aggregating them as described in Section~\ref{sec:method}, we analyze these webpage-visit level results to investigate 1) the prevalence of EIDs over time and 2) the characteristics of popular EID providers.

\subsection{Prevalence of EIDs}
\label{subsec:prevalence}

The HTTP requests in our dataset come from 17,028,370 crawler visits. On average, a single visit sends requests to 1.77 of the 6 SSPs, totaling 8.54 requests. Of the 616,539 unique websites, 72.34\% transmitted an EID during at least one visit.

Figure~\ref{fig:prevalence_eids_time}(a) plots the percentage of mobile/desktop visits resulting in at least one EID being transmitted over time. The figure also shows the percentage of websites with at least one EID transmitted across all visits in the given month. We observe that EID prevalence has greatly increased in the past few years.
In January 2022, only 28.92\% of visited websites resulted in an EID being transmitted across at least one visit to its webpages. Prevalence rose over the next year by about 10\%, before suddenly jumping to 47.83\% of websites in April 2023. This sudden increase seems explained by Amazon beginning to receive EIDs via client-side HTTP requests in April 2023 (see Appendix~\ref{app:ssp_eid_prevalence} for a full breakdown of how EID prevalence by SSP changed over time). EID prevalence was then stable for several months before again increasing nearly 18\% to 64.74\% of websites in September 2023. This jump occurred in the same month that Google announced that the Privacy Sandbox (i.e., a privacy-friendly alternative to third-party cookies) was generally available on Chrome and that cookies would begin to be deprecated for a subset of users in January 2024~\cite{sandbox_announcement_sept_2023}. 
By February 2024, one month after third-party cookie deprecation begins, 69.77\% of websites visited transmit at least one EID for header bidding. We believe this supports the idea that early increases in EID prevalence were largely driven by the digital advertising industry preparing for EIDs to replace third-party cookies. 

Following a July 2024 announcement that third-party cookie deprecation was being abandoned~\cite{sandbox_announcement_july_24}, the percentage of websites transmitting EIDs dropped 5\%. Although this continued to fall to 58.09\% in November 2024, the prevalence of EID unexpectedly began to increase again in December 2024. By the end of our data collection period, 83.76\% of visited websites transmitted at least one EID, an all-time high despite no future plans for third-party cookie deprecation on Google Chrome. This suggests that EIDs are no longer viewed solely as a third-party cookie replacement; this trend may be driven by the perceived advantages of EIDs over third-party cookies (e.g., cross-device tracking).

\begin{comment}
\begin{figure*}[h!]
  \centering
  \begin{subfigure}[b]{0.9\textwidth}
    \includegraphics[width=\textwidth]{images/EID_prevalence_time_1.pdf}
    \caption{Percentage with EIDs}
    \label{fig:prevalence_eids_time}
  \end{subfigure}

  \vspace{1em} %

  \begin{subfigure}[b]{0.9\textwidth}
    \includegraphics[width=\textwidth]{images/EID_prevalence_time_freq_2.pdf}
    \caption{Average Number of EIDs}
    \label{fig:avg_eids_time}
  \end{subfigure}

  \caption{\centering Prevalence of EIDs Over Time in Different Categories: (a) Percentage of Category with at least 1 EID (b) Average Number of EIDs per Webpage-Visit}
  \label{fig:combined_figures}
\end{figure*}

\end{comment}

Figure~\ref{fig:prevalence_eids_time}(b) presents a different perspective on EID prevalence: the average number of identity providers generating an EID for each visit over time. The overall trend again shows that EIDs are becoming more prevalent over time. However, there are two unexpected declines in the average number of transmitted EIDs that do not align with the trends in Figure~\ref{fig:prevalence_eids_time}(a): September 2023 and April 2025. Both of these sudden declines occur in months where the percentage of visits with at least one transmitted EID increases drastically. One explanation for this seemingly counterintuitive finding is that many of the new adopters of EIDs start by implementing EIDs from a single (or small number of) provider(s). Evidence in Appendix~\ref{sec:app_new_adopters} supports this explanation and suggests that these declines in Figure~\ref{fig:prevalence_eids_time}(b) should not be interpreted as evidence of declining EID prevalence. 

Figure~\ref{fig:prevalence_eids_time} also shows that EIDs have become more prevalent among both desktop and mobile visits, with similar trends in both subgroups. There is a small difference in terms of EIDs being more prevalent among desktop visits. In addition to Figure~\ref{fig:prevalence_eids_time}(a), we confirm this finding on the 4,556,372 webpage-date observations that have a mobile and desktop visit. Among them, EIDs are found in 53.48\% of the desktop visits and 49.39\% of the mobile visits. This difference may be due to EID generation and ID bridging being more difficult on mobile devices due to fewer browser log-ins (due to mobile apps) and less consistency in terms of the IP address used in a browsing session.

\begin{table}[b!]
\centering
\small
\begin{tabular}{|l|c|l|c|}
\hline
\textbf{Rank band} & \textbf{\%} &
\textbf{Rank band} & \textbf{\%} \\
\hline
Top 1K   & 73.79 & 50K--100K  & 77.17 \\
1K--5K   & 70.83 & 100K--500K & 80.15 \\
5K--10K  & 72.33 & 500K--1M   & 83.55 \\
10K--50K & 74.86 & Beyond 1M  & 84.26 \\
\hline
\end{tabular}
\caption{\centering Percentage of websites transmitting at least one EID, by CrUX popularity rank (May 2025).}
\label{tab:eid_by_rank}
\end{table}

\textbf{Prevalence by website popularity.} Because the HTTP Archive draws its crawl list from CrUX, every crawled website carries a popularity rank, which lets us determine whether EID adoption varies by popularity. Table~\ref{tab:eid_by_rank} reports EID prevalence across the websites in each CrUX rank group from May 2025, the final month of our observation period. With the exception of the top 1K websites having slightly higher EID usage than websites ranked in either the top 1K-5K or 5K-10K, EID prevalence has an inverse relationship with popularity. Websites outside of the top 1M are 14.2\% more likely to transmit an EID than the most popular websites. Despite this, all website popularity groupings show substantial EID usage with over 70\% of websites adopting in May 2025.

\subsection{EID Sharing Behaviors}
\label{subsec:provider_characteristics}

A second goal of our analysis is to examine the behavior of the popular identity providers. Although there are 731 different identity providers in our dataset, only 18 produce EIDs for at least 1\% of the webpage visits. These 18 providers account for 99.42\% of all transmitted EIDs in our dataset, and thus we focus our analyses on them (see Table~\ref{tab:eid_providers}). A deeper investigation into how EIDs are sent to SSPs helps to evaluate whether EIDs are capable of replacing third-party cookies as a privacy-preserving alternative tracking identifier. Several findings indicate potential accuracy and privacy concerns in the current implementation of EIDs.

\begin{table}[t!]
\centering
\resizebox{\columnwidth}{!}{%
\begin{tabular}{|c|cc|cc|}
\hline
\textbf{Identity} &          \multicolumn{2}{c|}{\textbf{1 SSP}} & \multicolumn{2}{c|}{\textbf{2+ SSPs}}\\
 \textbf{Provider}  & \textit{N} & \textit{2+ IDs (\%)} & \textit{N}  & \textit{2+ IDs (\%)} \\
\hline
\hline
 PubCID  & 4,975,548 & 2.12 & 3,538,489 & 20.25 \\
 ID5     & 1,958,855 & 8.14 & 1,519,459 & 32.35 \\
Criteo  & 1,678,856 & 14.73 & 720,531    & 20.71  \\
Trade Desk   & 1,260,865 & 0.12 & 774,017  & 0.28\\
 Audigent &  1,289,842 & 1.37 & 158,879   & 31.87 \\
 LiveIntent  & 341,182  & 15.64 & 488,058    & 23.60 \\
 Lotame &  616,991  & 0.05 & 177,376    & 0.08 \\
 33Across  & 296,883  & 0.88 & 406,401    & 14.01 \\
 Yahoo  &  315,020  & 0.01 & 317,142    & 2.52 \\
 Quantcast  &  528,527  & 0.22 & 55,808     & 0.82 \\
 Neustar & 80,333  & 4.64 & 447,978   & 6.88 \\
 Flashtalking   &  90,118   & 0.00 & 350,386   & 0.00 \\
 LinkedIn  & 217,931  & <0.01 & 104    & 0.00 \\
 Intimate Merger & 163,261  & 4.68 & 4,373   & 80.80 \\
 Pubmatic   & 159,566  & 0.03 & 5,520    & 0.29 \\
 Bid Switch  & 155,313  & 0.06 & 4,249      & 0.35 \\
 OpenX  & 126,366  & <0.01 & 5,076    & 0.00\\
 Rubicon& 109,352  & 0.04 & 4,837   & 0.27 \\
\hline
\hline
\textbf{Total} & \textbf{14,364,809} & \textbf{4.18} & \textbf{8,978,683} & \textbf{18.10} \\
\hline
\end{tabular}%
}
\caption{Multiple EIDs Sent by Popular EID Providers}
\label{tab:eid_providers}
\end{table}

\subsubsection{Multiple EIDs for a Single Visitor}
\label{subsub:multiple}

Our first observation is that a single webpage visit frequently resulted in multiple unique EIDs being transmitted by a single identity provider. Surprisingly, this occurred when an identity provider's EID is transmitted to either a single SSP or multiple SSPs. In Table~\ref{tab:eid_providers}, we report the percentage of website visits resulting in multiple unique EIDs transmitted by each identity provider. We break this down by whether the EID goes to one SSP or several (``1 SSP'' and ``2+ SSPs'').

When an identity provider transmits an EID to a single SSP, multiple unique EIDs are sent by the provider in 4.18\% of cases. For identity providers where transmitting multiple EIDs to a single SSP is rare (e.g., OpenX), such occurrences may indicate errors by the identity provider or the website. For providers where it is more common (e.g., ID5) and cannot be attributed to random malfunctions, there are two options: 1) all EIDs resolve to the same user, or 2) the EIDs resolve to two different users. Although seemingly unnecessary, it is only an inconvenience if the EIDs do resolve to the same user. However, if the EIDs resolve to different users, it raises serious questions about the accuracy of EIDs. Unfortunately, without decryption access, we cannot confirm how frequently the different transmitted EIDs resolve to the same user. 

Identity providers are more likely to send multiple EIDs during a single webpage visit  (18.1\% of the time) when they are transmitting EIDs to multiple SSPs. Again, this issue is most problematic if different EIDs are not all resolving to the same user identity. In the setting where multiple SSPs are involved, this would be an even larger issue, as SSPs pass the EIDs on to advertisers who are bidding against each other to serve advertisements. Different resolved identities across SSPs raise fairness concerns.

One particularly concerning observation is that most identity providers do not appear to follow a uniform strategy. Only three providers (Flashtalking, LinkedIn, and OpenX) always transmit one EID to all SSPs, and no provider always transmits unique EIDs to each SSP. Even when multiple distinct EIDs are transmitted, 41.99\% of the time a unique EID was transmitted to every participating SSP, and 58.01\% of the time, one of the EIDs was transmitted to at least two SSPs. The lack of a consistent strategy is alarming because we would expect to see consistency (e.g., always transmitting one unique EID per SSP or always transmitting a single EID to all SSPs) if the generation of multiple EIDs was a privacy-preserving choice rather than a consequence of uncertain identity resolution.

\subsubsection{Bot Detection}
\label{subsub:bots}

A serious issue we uncover is that many identity providers fail to list the crawler's identity as a bot, despite the HTTP Archive self-declaring its crawler as bot traffic. Flashtalking is the only identity provider in our dataset to explicitly identify the crawler as a bot (i.e., the EID value is ``Bot''). Other identity providers infrequently assign null EIDs (e.g., ``0'' or ``null''), but it is unlikely this has to do with bot recognition. For example, ID5, which is the only company assigning null identities more than 0.25\% of the time, states that the null identity (i.e., ``0'') indicates that the user did not consent~\cite{id5prebid}. Assigning EIDs to bot traffic could harm advertisers. If the identity corresponds to a known user, advertisers may falsely believe they are reaching that user. Even if it does not resolve to a known identity, the presence of an EID alone may lead advertisers to believe the impression is served to a real user. Either way, advertisers pay to show ads to a bot. 

\subsubsection{Reusing Identifiers.}\label{subsub:reuse} EIDs are supposed to be more privacy-preserving partly because the encrypted identifier transmitted to SSPs is updated for every visit, so no third party can use it to track a user across websites~\cite{sivan2025}. If the same EID is transmitted across multiple webpage visits from a user, third-party trackers embedded in a first-party context~\cite{munir2023cookiegraph,chen2021cookieswap} or using CNAME cloaking~\cite{munir2023cookiegraph,dimova2021cname,dao_cname} may access the EIDs stored as first-party cookies for tracking. We find that 12 of the 18 providers do not always generate new EIDs for each webpage visit. Across our dataset, there are 151,900 persistent EIDs (i.e., non-null identifiers transmitted across multiple webpage visits). Table~\ref{tab:persistence_eids_agg} reports their average persistence, and we break it down by provider in Appendix~\ref{app:persistence}. 

\begin{table}[t]
\centering
\small
\caption{Persistence of Reused EIDs}
\begin{tabular}{lcc}
\hline
\textbf{Statistic} & \textbf{Average} & \textbf{Max}\\
\hline
Number of Webpage Visits & 20.59 & 57,366 \\
Number of Distinct Websites & 19.54 & 36,529 \\
Months & 1.35 & 8 \\
\% on Desktop and Mobile Visits& 26.82 & - \\
\hline
\end{tabular}
\label{tab:persistence_eids_agg}
\end{table}

On average, the persistent EIDs identified in our dataset were transmitted across 20.59 different webpage visits to 19.54 unique websites over 1.35 months (only 13.19\% of persistent EIDs were transmitted across multiple months). This indicates that the average persistent EID could be used to track a user across 20 different webpage visits. If persistent EIDs from multiple providers are available, these could be synced by trackers to enable larger-scale cross-site tracking. Furthermore, in worst-case scenarios, EIDs were extremely persistent, including one example where an EID from Audigent was transmitted across 52,046 distinct crawls to 29,770 unique websites from November 2024 to January 2025. We display the full distribution of EID persistence in Appendix~\ref{app:eid_frequency}.

Persistent EIDs are not only a privacy concern. In total, 26.82\% of the persistent EIDs were used to represent both mobile and desktop visitors. The HTTP Archive's methodology~\cite{httparchive_method} prevents mobile and desktop crawls from being linked by user log-ins or third-party cookies. The devices share no characteristics, other than potentially using the same IP address (which is selected from a rotating list of Google Cloud IPs). Determining user identity solely based on IP address, especially when those IP addresses are the commonly provided IP addresses from Google Cloud, is not likely to produce meaningful identifiers.

\subsubsection{Do EID providers respect user consent?}\label{subsub:consent} 

The idea that EIDs are more privacy-preserving than cookies is based on the fact that the identity provider has control over whether or not an EID is transmitted (and to whom)~\cite{id5id, iiq_EIDs, audigent2023hadron, admonsters_eids_10, sivan2025}. The HTTP Archive crawler does not interact with any cookie banners that may ask for user consent~\cite{httparchivecookies}, and yet we found 34,700,613 transmitted non-null EIDs in our dataset. The lack of strict privacy regulation in many countries (e.g., the United States) may allow identity providers to generate and transmit EIDs unless the user explicitly asks them to stop or visits their website and opts out. In Appendix~\ref{app:id5}, we present evidence suggesting that ID5, a major EID provider, may have internally changed their policy to stop requiring explicit opt-in user consent before transmitting an EID. 

In theory, practices should be different in the EU where GDPR requires explicit user consent \emph{before} the user's personal data is collected or processed. 
We use each website's country code top-level domain (ccTLD) as a \emph{proxy} for whether it is operated from a GDPR-protected country (i.e., the EU or UK). Of the 52,475 identified websites, a non-null EID is generated at least once on 62.67\%. Only Flashtalking and Intimate Merger generated no EIDs on websites with ccTLDs corresponding to a GDPR-protected country.

\section{Discussion and Conclusions}
\label{conclusion}

This study is the first analysis of EID adoption on the Web. Our findings show that EID adoption has increased substantially in the last few years. Surprisingly, adoption did not abate after Google abandoned third-party cookie deprecation in July 2024~\cite{sandbox_announcement_july_24}. This suggests that the digital advertising industry views EIDs as more than a third-party cookie substitute, possibly due to supposed benefits such as cross-device tracking and privacy-preserving design.

However, our analysis of 18 popular identity providers suggests EIDs are not yet mature as a trustworthy alternative identity solution. First, identity providers show inconsistent behavior when transmitting EIDs, frequently transmitting multiple EIDs for a single visitor or sending different EIDs to competing SSPs. This may introduce inaccuracies and unfairness into auctions as advertisers are competing against each other with different assumptions about the targeted user. Second, we found that 17 of the 18 providers failed to identify the self-reported crawler as bot traffic, meaning that advertisers may unknowingly pay to serve ads to bots. With increasing bot traffic due to AI training, failure to properly identify this traffic may become a substantial problem.

Finally, we question whether EIDs truly preserve user privacy as claimed. Without any user consent, 16 of the 18 providers transmitted EIDs when the crawler visited EU websites, possibly violating GDPR's explicit consent requirement. Furthermore, 12 providers transmit the same EID across multiple visits, websites, and devices. Persistent EIDs may allow third-party trackers to perform cross-site or cross-device tracking without the identity providers' knowledge, which circumvents the privacy-preserving design of EIDs.
These problems harm advertisers and users alike. Inaccurate identifiers degrade the retargeting and measurement advertisers pay for, while identifiers generated without consent and reused across visits, websites, and devices threaten user privacy. %

\textbf{Limitations.} Our findings can only definitively be interpreted as reflecting how EIDs work when an automated crawler visits
websites in a simulated Chrome browser. Given that EIDs were assigned to the self-declared crawler, we have no reason to believe the findings are not representative of how EIDs are assigned to real users. Even though we cannot decode the values, the mix of persistent and unique IDs suggests that these are neither random nor default values. Future research should evaluate EIDs originating from real-user visits, in browsers that restrict third-party cookies and fingerprinting (e.g., Safari), and in response to particular user behaviors (e.g., after logging into a user account). 

Furthermore, our data collection limits our analyses to the EIDs that we could manually extract in client-side header bidding requests sent to six SSPs that frequently participated in these auctions. EIDs may be used differently by other SSPs, or in other contexts (e.g., when SSPs receive server-side bid requests or when publishers sell directly to advertisers). Future research should examine EID use across other methods of selling ad inventory, including direct-sold or server-side programmatic auctions.

Given the increasing prevalence of EIDs and the highlighted issues, we call for future research and industry standards to increase transparency and trust in the implementation of EIDs. Specifically, we advocate for industry standards that require identity providers to state what identity signals are used. Advertisers should know, and be able to bid differently, when EIDs are generated from deterministic versus probabilistic ID bridging techniques. Audits of ID bridging technologies should be required to ensure accuracy for advertisers and privacy for users who have opted out. Finally, it is important that EIDs comply with all relevant privacy regulations. Given differing regulations across the globe, one solution is to shift responsibility to websites, invoking the identity provider only when compliance requirements are met. Websites already comply with privacy regulations elsewhere, so this may increase compliance. %

\section{Acknowledgements}
\label{sec:acknowledgements}

This work was supported in part by the National Science Foundation under grants CNS-2237328 and DGE-2623258, by the National Center For Advancing Translational Sciences of the National Institutes of Health under Award Number UM1TR004789, by the Martin Tuchman '62 Chair Endowment, and by the Leir Foundation.

\bibliographystyle{ACM-Reference-Format}
\bibliography{references}

\appendix
\raggedbottom
\begin{figure*}[tp]
\includegraphics[width=0.85\textwidth]{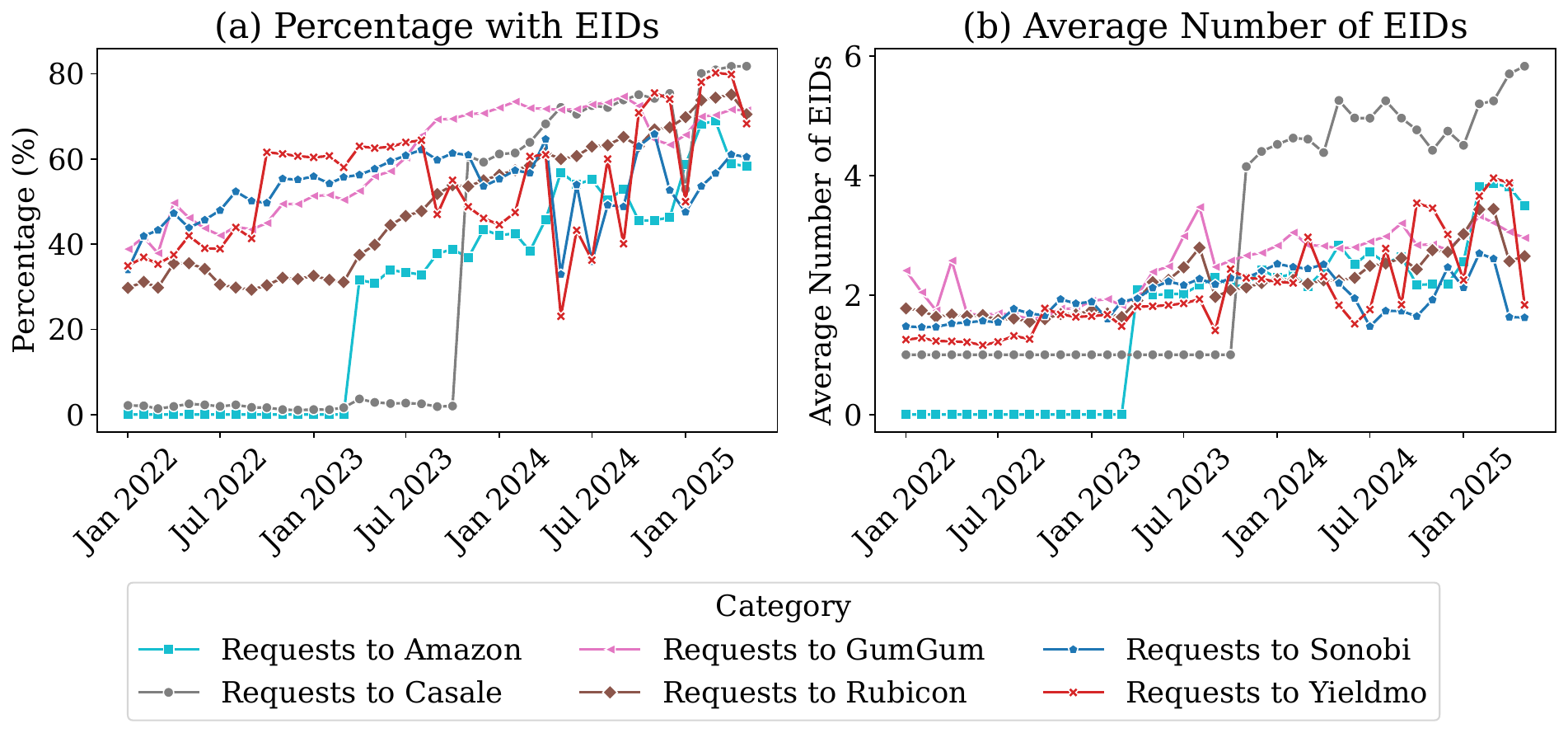}
    \caption{\centering Prevalence of EIDs Over Time by SSP: (a) Percentage of Requests with at Least One EID, (b) Average Number of Distinct EIDs per Request}
    \label{fig:app_eids_time}
\end{figure*}

\section{Ethics}

The insights from this work have revealed several limitations in the current implementation of EIDs in the header bidding ecosystem. It is not the authors' intention that this study be used as a justification for third-party cookies. We present these findings to highlight issues that could affect industry stakeholders and suggest that the current implementation of EIDs is not as privacy-preserving as claimed by identity providers. It is important to establish trust through increased transparency and regulation so that EIDs can better serve the interests of stakeholders in digital media and the public. 

Although our analysis discusses privacy concerns and the assignment of EIDs without user consent, the analyzed dataset contains only self-identified crawler traffic, and user identifiers do not correspond to real individuals, minimizing privacy risks. 

\section{EID Prevalence by SSP}
\label{app:ssp_eid_prevalence}

Figure~\ref{fig:app_eids_time} shows the prevalence of EIDs over time for each of the six observed SSPs. Although prevalence in terms of the percentage of requests sent to an SSP containing at least one EID (Figure~\ref{fig:app_eids_time}(a)), and the average number of distinct EIDs transmitted in a request (Figure~\ref{fig:app_eids_time}(b)) both increase over the time period of our dataset, this trend includes much more noise than the trend aggregated to visit or website level (Figure~\ref{fig:prevalence_eids_time}). Some of this noise is a result of multiple HTTP requests being sent to a single SSP from a single crawler visit, where EIDs are only included in some of the HTTP requests. 

\section{Relationship Between New Adopters and Average Number of EIDs}
\label{sec:app_new_adopters}

A possible explanation for the unexpected declines in the average number of EID providers employed by a publisher in September 2023 and April 2025 is that EIDs were introduced to many new websites. Because the average number of EID providers used by each publisher in Figure~\ref{fig:prevalence_eids_time}(b) only includes publishers using at least one EID provider, it is susceptible to a surge in new publishers adopting EIDs from a small number of EID providers. Intuitively, publishers may only work with 1 or 2 EID providers as part of a testing phase to better understand the value of EIDs. 

Our descriptive results suggest that this interpretation may be accurate. In the two months (i.e., September 2023 and April 2025) where the average number of EIDs declined, the newly adopting websites utilized just 1.69 and 1.71 EID providers on average, respectively. The average number of EID providers utilized among all websites in those months was 2.21 and 3.05, respectively. 

Furthermore, the decline in the average number of transmitted EIDs in September 2023 was only short term. The average number of EID providers utilized began to rise again the following month, and by May 2024, the average number of unique providers' EIDs transmitted per crawl had reached a new all-time high in our dataset. Visits to the websites that had just adopted EIDs in September 2023 utilized EIDs from 2.81 providers on average by May 2024, representing a 64\% increase in the average number of EIDs transmitted per crawl in comparison to September 2023. 

\section{Persistent Identifiers by Identity Providers}
\label{app:persistence}
\begin{table*}[ht]
\centering
\small
\begin{tabular}{|c|cc|ccc|cc|}
\hline
\makecell{\textbf{Identity} \\ \textbf{Provider}} & \makecell{\textbf{Number of} \\ \textbf{Persistent IDs}}  & \makecell{\textbf{\% of} \\ \textbf{Total IDs}} & \makecell{\textbf{Avg \# of} \\ \textbf{Unique} \\ \textbf{Visits}} & \makecell{\textbf{Avg \# of} \\ \textbf{Unique} \\ \textbf{Websites}} & \makecell{\textbf{Max \# of} \\ \textbf{Unique} \\ \textbf{Websites}} & \makecell{\textbf{\% of Persistent} \\ \textbf{IDs Used}\\ \textbf{Across Months}}& \makecell{\textbf{\% of Persistent} \\ \textbf{IDs Used}\\ \textbf{Across Devices}} \\
\hline
Audigent          & 71,263 &  63.94 & 20.73 & 18.90  & 36,259  & 28.05 & 29.63\\
 33Across          & 44,713  & 6.30  & 2.22 & 2.22  & 18  & 0.00 & 0.00\\
 Lotame            & 9,123  &  2.32  & 44.76 & 44.71   & 1,617  & 0.08 &2.49\\
  LiveIntent        & 8,799  & 1.33  & 45.88 & 45.52  & 835  & 0.19 & 84.99\\
 Pubmatic      & 4,636 & 80.12  & 35.38 & 35.32    & 361  & 0.06 & 84.69 \\
 Intimate Merger    & 3,873  &24.25 & 43.04  & 36.80     & 4,419 & 0.36 & 0.36\\
 Bid Switch    & 3,860  & 79.97  & 41.11& 41.01    & 401  & 0.05& 84.02\\
 OpenX            & 2,859  & 74.38  & 45.63 & 45.53    & 452  & 0.03& 84.78\\
 Rubicon & 2,511  & 72.87  & 45.13 & 45.05    & 441  & 0.16& 83.63\\
Neustar       & 258    & 0.05  & 25.42 & 25.02 & 420  &0.00 & 81.01\\
Yahoo             & 3  &  <0.01  & 2.00 & 2.00  & 2  &  0.00 & 33.33\\
Trade Desk              & 2  & <0.01  & 35.00 & 2.50  & 4  & 100.00 & 50.00\\
\hline
\end{tabular}
\caption{Identity Providers with Persistent EIDs}
\label{tab:persistent_ids}
\end{table*}

Table~\ref{tab:persistent_ids} shows that Audigent transmits the most persistent EIDs (71,263), and also that persistent EIDs represent a large percentage of its total transmitted EIDs (63.94\%). No provider transmits only persistent EIDs, and three of the twelve providers (Neustar, Yahoo, and the Trade Desk) transmit very few persistent EIDs. The Trade Desk specifically represents an odd case where it only transmits two persistent EIDs (less than 0.0001\% of their total transmitted EIDs), and yet one of the persistent EIDs is highly persistent as it is transmitted on 68 different crawls spanning 8 months. 

\section{Distribution of EID Occurrences}
\label{app:eid_frequency}

\begin{figure}[t]
\centering
\includegraphics[width=\linewidth]{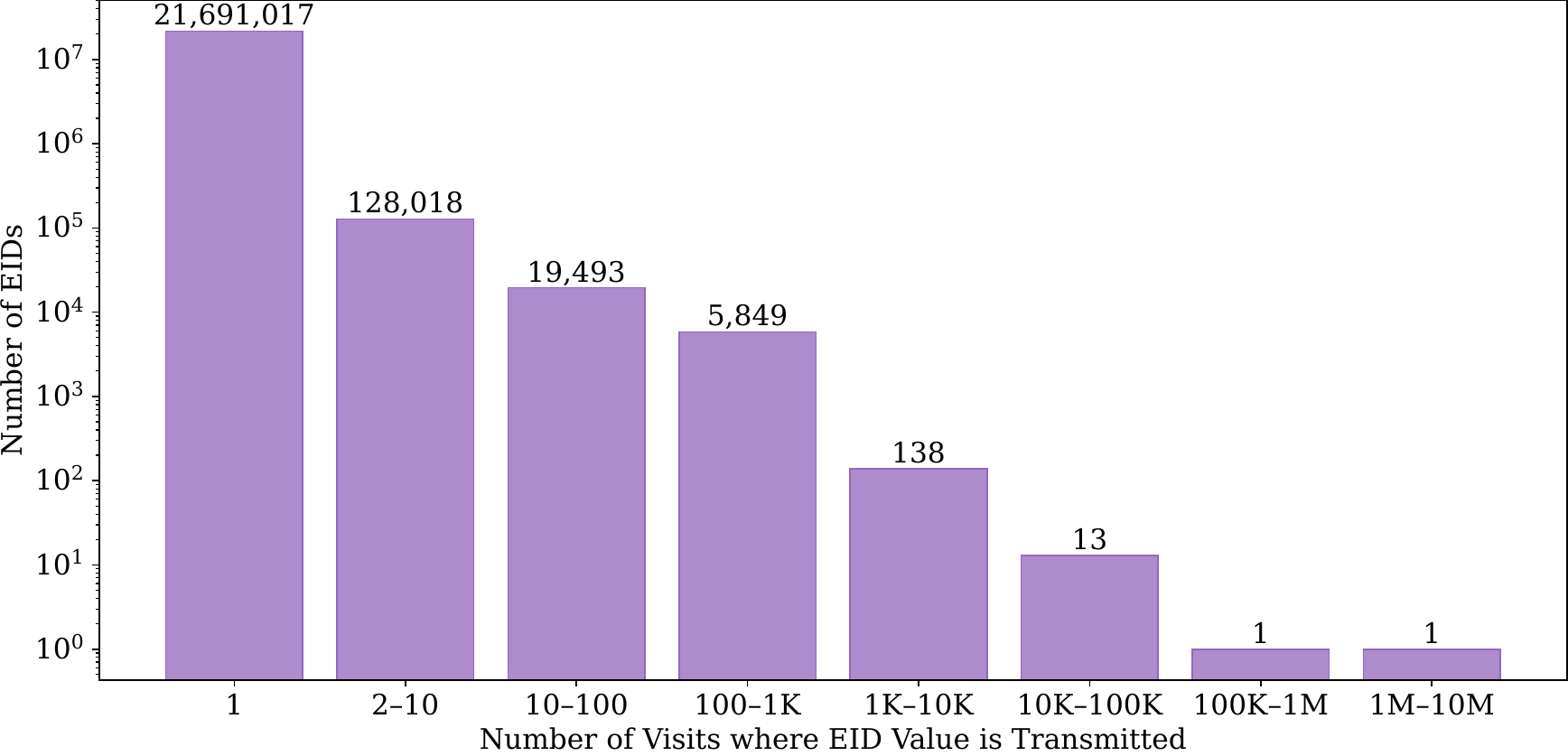}
    \caption{\centering Number of distinct EID values by how many webpage visits transmit them. Repeated transmissions of the same value within a single visit are counted once. Note the logarithmic $y$-axis.}
    \label{fig:eid_value_frequency}
\end{figure}

Figure~\ref{fig:eid_value_frequency} shows the distribution for how many distinct webpage visits each EID in our dataset is transmitted on. The vast majority of EID values, 21,691,017 of them, are transmitted on exactly one webpage visit, which is what a privacy-preserving design should produce. An identifier that is regenerated per visit cannot be used by an unintended third party to link visits together.

Only two values in the dataset are transmitted more than 100,000 times, and neither is an actual identifier. The most frequent is ID5's ``0'', transmitted 1,503,865 times during the period in which ID5's internal policy did not permit it to transmit an EID without explicit consent (see Appendix~\ref{app:id5}). The second is Flashtalking's ``Bot'' value, transmitted 440,504 times. Flashtalking is the one provider in our dataset that explicitly identifies our self-declared crawler as automated traffic (Section~\ref{subsub:bots}). Both are documented  default values rather than user identifiers.

This distribution is the empirical basis for the argument in the Limitations paragraph of Section~\ref{conclusion} that the EIDs we observe are neither random strings nor default values. If providers were transmitting default values, we would expect a small number of values with very high frequency for each provider; instead only the two documented placeholders above exceed 100,000 occurrences. If providers were emitting random strings, there would be no explanation for the values that occur across separate, stateless webpage visits, which are the persistent EIDs analyzed in Section~\ref{subsub:reuse}.

\section{Extracting EIDs from Header Bidding Requests}
\label{app:extraction}

This appendix expands on the EID extraction procedure described in Section~\ref{sec:method}. Within the request, EIDs are stored together in a unified format. Once the formatting has been determined within a request, recovering the transmitted EIDs (with the identity provider that generated them) does not require any inference.

The more involved step is determining the format, because the six SSPs do not encode their incoming header bidding requests the same way. Some SSPs transmit EIDs as individual URL query parameters, while others embed EIDs within a JSON request body or JSON objects encoded as query parameter values. For each SSP we determined the encoding by combining two sources: the publicly available Prebid adapter code for that SSP (including all changes over the time period studied), which documents how the request is assembled~\cite{prebid_github}, and manual inspection of raw requests to each SSP in our dataset. We then implemented one parser per SSP, so that a request is only searched in the location where that SSP is known to place EIDs. Table~\ref{tab:ssp_encoding} summarizes where each SSP places the EID list.

\begin{table}[h!]
\centering
\caption{Location of the EID list in each SSP's header bidding request}
\begin{tabular}{lp{0.55\columnwidth}}
\hline
\textbf{SSP} & \textbf{Request encoding} \\
\hline
Magnite (Rubicon) & Query string, one \texttt{eid\_<source>} parameter per provider \\
Amazon Ads & JSON inside URL query parameter\\
Index Exchange (Casale) & OpenRTB JSON in the \texttt{r} URL parameter, at \texttt{user.eids} \\
Sonobi & JSON array in the \texttt{eids} query parameter \\
GumGum & Query string, one parameter per provider keyed by identifier name \\
Yieldmo & JSON array in the \texttt{eids} query parameter \\
\hline
\end{tabular}
\label{tab:ssp_encoding}
\end{table}

Listings~\ref{lst:ex_magnite}--\ref{lst:ex_yieldmo} show a request to each SSP with the EID list highlighted. Publisher and account identifiers are replaced with \texttt{<REDACTED>}, identifier values are truncated, and long lists are elided; the encoding is otherwise unmodified. The same logical structure appears three ways: one query parameter per provider at Magnite and GumGum, a JSON array in a single \texttt{eids} parameter at Sonobi and Yieldmo, and nesting at differing depths inside a JSON object at Amazon Ads and Index Exchange. Magnite and GumGum differ again, since Magnite keys by source domain and GumGum by the provider's product name.

\begin{listing}[h!]
\begin{lstlisting}[style=eidexample]
https://fastlane.rubiconproject.com/a/api/fastlane.json
  ?account_id=<REDACTED>&site_id=<REDACTED>&zone_id=<REDACTED>&...
@@  &eid_id5-sync.com=ID5*Hcje60BnAY_ZhXdB...
  &eid_pubcid.org=fba40ec8-ec72-4e21-b...
  &eid_criteo.com=_NdwmF9pcU5wdEFkcUd0...@@
  &rf=<page URL>&tk_flint=pbjs_lite_v8.51.0&rand=...
\end{lstlisting}
\caption{Magnite. EIDs arrive as one \texttt{eid\_<source>} query parameter per provider, so the provider name is carried in the parameter name rather than in a \texttt{source} field.}
\label{lst:ex_magnite}
\end{listing}

\begin{listing}[h!]
\begin{lstlisting}[style=eidexample]
https://aax.amazon-adsystem.com/e/dtb/bid?...
@@  &vm={"ids":{"audigent":"060f87bcgl...",
    "id5":"ID5*8NKfZa19Q....",
    "lotame":"59a93bb34868.....",
    "pubcommon":"6acc6b06-1b5..."},
@@ "vendors"...}... 
\end{lstlisting}
\caption{Amazon Ads. The list \texttt{ids} sits inside the JSON in the \texttt{vm} parameter.}
\label{lst:ex_amazon}
\end{listing}

\begin{listing}[h!]
\begin{lstlisting}[style=eidexample]
https://as-sec.casalemedia.com/cygnus?<REDACTED>
  ...&r={"id":"...", "imp":[...], "site":{...},
@@  "user": {"eids": [
    { "source": "criteo.com",
      "uids": [ { "id": "6vuJgV83eUI2MFdU...", "atype": 1 } ] },
    ... (4 further providers omitted)
  ]@@
  }} ... &...
\end{lstlisting}
\caption{Index Exchange. The list sits at \texttt{user.eids} inside the JSON in the \texttt{r} parameter. }
\label{lst:ex_indexexchange}
\end{listing}

\begin{listing}[h!]
\begin{lstlisting}[style=eidexample]
https://apex.go.sonobi.com/trinity.json?<REDACTED>&...
@@&eids=[
  { "source": "criteo.com",
    "uids": [ { "id": "VoDdMl9XUUpvZXJH...", "atype": 1 } ] },
  { "source": "id5-sync.com",
    "uids": [ { "id": "0", "atype": 1, "ext": {"linkType": 0} } ] },
  { "source": "adserver.org",
    "uids": [ { "id": "3f66c98a-619d-4a...", "atype": 1, "ext": {"rtiPartner": "TDID"} } ] },
    ... (2 further providers omitted)
]@@
&...
\end{lstlisting}
\caption{Sonobi. A JSON array in the \texttt{eids} query parameter. The \texttt{id5-sync.com} value is the null placeholder \texttt{0}.}
\label{lst:ex_sonobi}
\end{listing}

\begin{listing}[h!]
\begin{lstlisting}[style=eidexample]
https://g2.gumgum.com/hbid/imp?si=<REDACTED>&pi=3&bf=728x90,728x250
  &pu=<page URL>&vw=1920&vh=1007&dpr=1&...
@@  &pubcid=c37f716b-6aee-4c9d-801...
  &tdid=b25c67d6-78cc-41e5-946...
  &id5id=0
  &criteoId=z1zGSF8xU1NaMXFjZWFwbE...
  &fabrickId=E1:2GcjVj6AH727kWwmruB...
  &lotamePanoramaId=9eee23431f32690d68f96d...@@
  &schain=1.0,1!<REDACTED>&uspConsent=1YNN&ns=4864
\end{lstlisting}
\caption{GumGum. One query parameter per provider, keyed by the provider's own product name rather than by a \texttt{source} field. The most recent list of providers' keys can be found at \url{https://docs.prebid.org/dev-docs/modules/userId.html}.} 
\label{lst:ex_gumgum}
\end{listing}

\begin{listing}[h!]
\begin{lstlisting}[style=eidexample]
https://ads.yieldmo.com/exchange/prebid?<REDACTED>&...
@@&eids=[
  { "source": "pubcid.org",
    "uids": [ { "id": "6d4d7804-cf02-43...", "atype": 1 } ] }
]@@
&...
\end{lstlisting}
\caption{Yieldmo. A JSON array in the \texttt{eids} query parameter (same format as Sonobi), here carrying a single identifier.}
\label{lst:ex_yieldmo}
\end{listing}

\section{Robustness to Changes in the Crawled Website List}
\label{app:subset}

\begin{figure}[t]
\centering
\includegraphics[width=\linewidth]{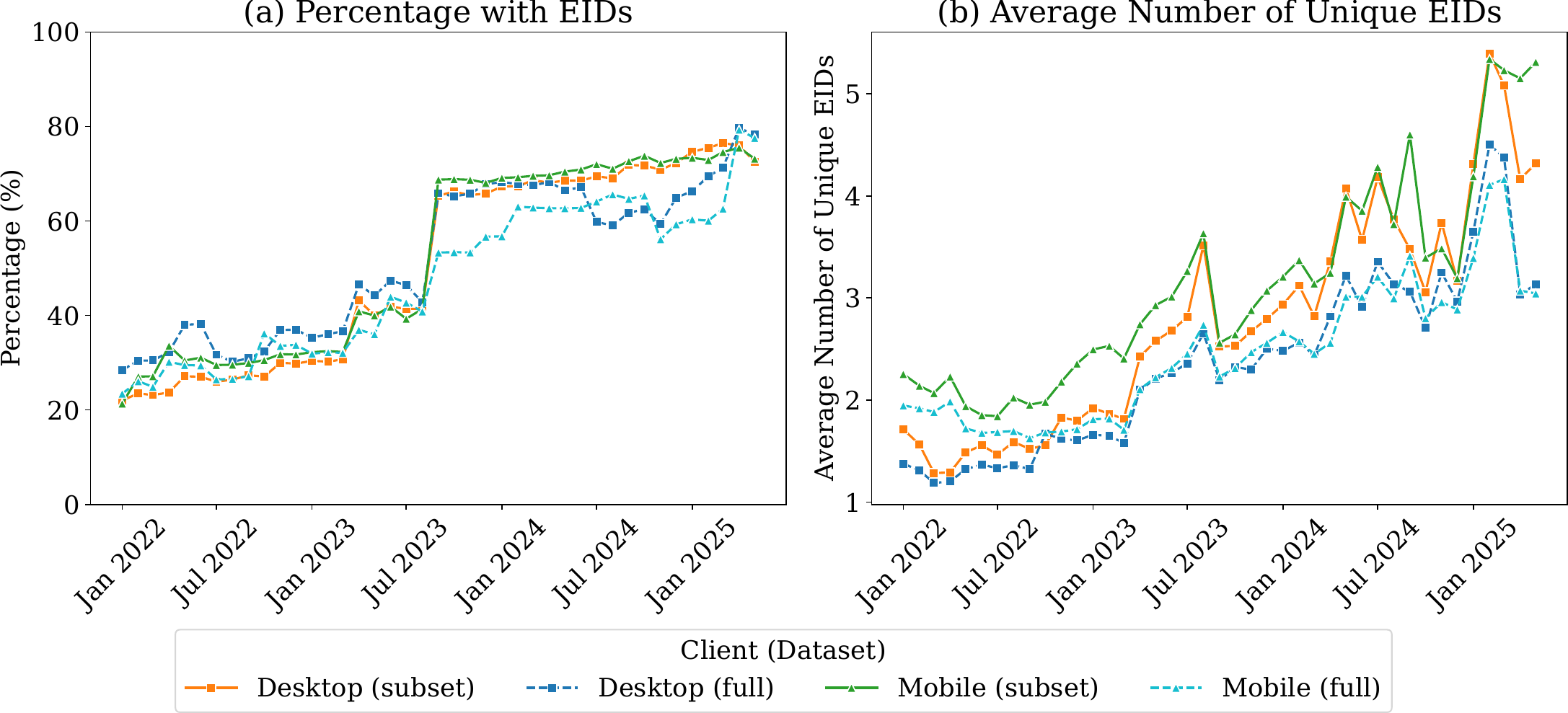}
    \caption{Prevalence of EIDs over time in the full dataset and in the fixed subset of websites crawled in both January 2022 and May 2025: (a) percentage of visits transmitting at least one EID, (b) average number of distinct EIDs per visit.}
    \label{fig:prevalence_subset}
\end{figure}

Because the HTTP Archive selects its crawl list from the Chrome UX Report, that list changes over the 41 months of our study (Section~\ref{sec:method}), so the growth reported in Section~\ref{subsec:prevalence} could reflect a change in composition rather than in adoption. We therefore restrict the analysis to the 24,881 desktop and 26,321 mobile websites present in both the first and last months of the observation window to create a consistent subset for analysis.

Figure~\ref{fig:prevalence_subset} plots both panels of Figure~\ref{fig:prevalence_eids_time} for the subset alongside the full dataset, and our conclusions hold. Within the subset, the percentage of visits transmitting at least one EID rises from 21.76\% to 77.00\% on desktop (21.06\% to 76.92\% on mobile), and the average number of EIDs per visit increases from 1.71 to 4.32 on desktop (2.25 to 5.31 on mobile). In comparison to our main results in Section~\ref{subsec:prevalence}, this subset shows a smaller initial decline in usage post third-party cookie deprecation abandonment and a higher average EID count, consistent with these being larger, more established properties that work with more identity partners and were less likely to treat EIDs purely as a cookie substitute.

\section{Case Study of ID5 Policy Change Allowing EIDs to be Transmitted without Consent}
\label{app:id5}

\begin{figure}[H]
\includegraphics[width=\linewidth]{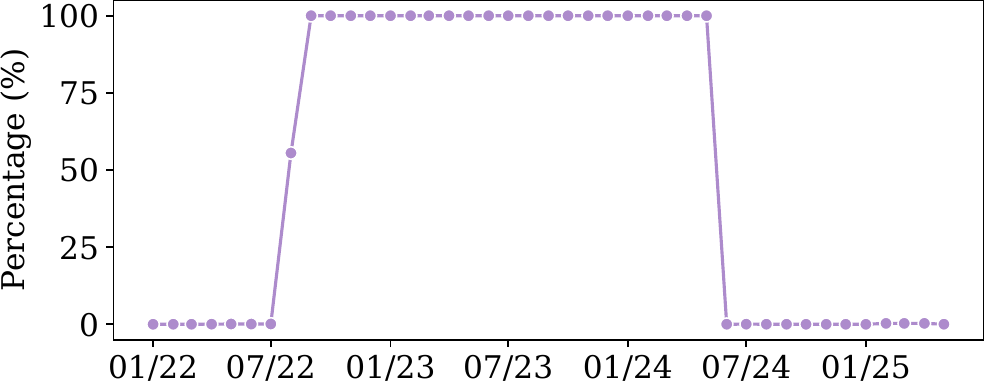}
    \caption{ID5 Null EIDs}
    \label{fig:ID5_null_EIDs}
\end{figure}

Figure~\ref{fig:ID5_null_EIDs} plots the percentage of webpage visits where ID5 transmits only null EIDs by month. ID5 transmits almost exclusively non-null EIDs at the start of 2022 (99.98\%), then shifts in August 2022 to generating null EIDs 100\% of the time by September. This continues until June 2024, when ID5 returns to transmitting non-null IDs almost exclusively (99.97\%), and is still doing so at the end of our dataset despite no explicit user consent. 

It is unclear why ID5 began requiring explicit consent in August 2022, or why it reversed that decision in June 2024. The change was, however, applied uniformly across all websites regardless of location, with implications for Web privacy. 
\end{document}